\documentclass[aps,prb,floatfix]{revtex4}
\usepackage{graphicx}

\begin{document}

\title{On the question of ferromagnetism in alkali metal thin films}
\author{Prasenjit Sen}

\affiliation{Harish-Chandra Research Institute, Chhatnag Road, Jhunsi, Allahabad 211019, INDIA.}

\begin{abstract}
Electronic and magnetic structure of $(100)$ films of K and Cs are calculated
within the plane-wave projector augmented wave (PAW) formalism of the density 
functional theory (DFT) using both local spin density approximation (LSDA)
and the PW91 generalized gradient approximation (GGA). Only a 6 layer Cs film
is found to have a ferromagnetic (FM) state which is degenerate with a
paramagnetic (PM) state within the accuracy of these calculations. This is at variance
with the results obtained from a finite thickness uniform jellium model (UJM).
Implications of these results for the experiments on transition metal doped alkali metal
thin films and bulk hosts are discussed.  

\end{abstract}

\maketitle

\section{Introduction}
Bulk and thin films of alkali metals containing 3$d$ transition metal (TM) 
impurities have been studied experimentally for quite some time now~\cite{riegel,kowallik,
beckmann,gambardella,bergsong}. The most intriguing
property of these systems is the large magnetic moment they posses. Beckmann
and Bergmann~\cite{beckmann}, through anomalous Hall measurements, show that
a Co impurity on a Cs film has a (total) magnetic moment as large as 9 $\mu_B$,
and that in bulk Cs has a moment of 8 $\mu_B$. An Fe impurity has a moment of 7
$\mu_B$ both on a Cs film, and inside bulk Cs. Bergmann and Song~\cite{bergsong}
predicted large moments in V doped alkali metals which ranged from 6-7 $\mu_B$
in a Na host to 4 $\mu_B$ in a Cs host. Nearly two decades ago, 
Riegel {\it et al}.~\cite{riegel} and Kowallik {\it et al}.~\cite{kowallik}, 
through magnetic susceptibility measurements,
showed that Fe and Ni impurities in K, Rb and Cs hosts have large magnetic
moments. Gambardella {\it et al.}~\cite{gambardella} in their X-ray absorption spectroscopy (XAS) and
X-ray magnetic circular dichroism (XMCD) measurements find total magnetic moments of 
6.63 $\mu_B$, 5.59 $\mu_B$, and 3.55 $\mu_B$ respectively for Fe, Co and Ni 
impurities on K films. 

These startling findings raise the obvious question of the origin of such large
magnetic moments. And that is where opinions differ. Riegel and co-workers~\cite{riegel} argued
that the TM atoms remain isolated inside an alkali host due to the latter's rather
large lattice constant. This isolated nature of the TM atoms retain their atomic
spin and orbital moments, which are, otherwise, quenched in a solid environment due
to hybridization and crystal-field effects. The moments are, therefore, essentially
those of isolated TM atoms. Gambardella {\it et al.}~\cite{gambardella} also subscribe to this view.
Comparing their observed XAS spectra with the calculated ones, they show that the
TM atoms are indeed in their atomic configurations. Beckmann and Bergmann~\cite{beckmann}
and Bergmann and Song~\cite{bergsong}, on the other hand have argued that the only
way to explain the observed moments in the TM-alkali metal systems is to assume
that the electrons in the alkali host are polarized in presence of the TM atoms. This
is exactly how an Fe impurity behaves in a Pd host, where it polarizes the Pd electrons
leading to a giant moment~\cite{crangle,bergmann}.

TM-alkali metal systems have been studied theoretically also. And the picture
that emerges supports Riegel {\it et al}. and Gambardella {\it et al}.'s views.
Through local density approximation (LDA) plus local Coulomb energy ($U$) calculations 
within the density functional (DFT) theory, Kwon and Min~\cite{kwon} 
have shown that an Fe impurity behaves differently in a Cs host compared to a Pd host.
There is no polarization of the Cs electrons, rather the Fe impurity remains isolated
leading to atomic-like large moments. Sahu and Kleinman~\cite{sahu} performed
DFT calculations for V impurities in bulk Na, and Co in Na and K. They
also rejected the hypothesis that polarization of the host electrons is
responsible for the observed moment.

In support of their proposition, Bergmann and co-workers cited Okazaki and
Teraoka's~\cite{okazaki1} work. In this, the authors use a uniform jellium model (UJM)
of finite thickness to study electronic and magnetic properties of alkali metal thin films. 
Using local spin density approximation (LSDA) for the exchange-correlation functional, 
they find that depending
on the electron density $\rho_0$ (equivalently $r_s=\left ( 3/4 \pi \rho_0\right )^{1/3}$), 
the UJM can have a ferromagnetic ground state in certain ranges of thickness ($D$).
In a later detailed work~\cite{okazaki2}, they show that for $r_s=6$, 
paramagnetic (PM) and fully polarized ferromagnetic (FPF) solutions exist at every thickness. 
They also find partially polarized ferromagnetic (PPF) solutions for $11.9 \le D \le 18.48$, 
and $22.2 \le D \le 27.6$ Bohr, and antiferromagnetic (AF) solutions for $11.5 \le D \le 15.0$ Bohr. 
The ground state of the system turns out to be FPF at $D \le 9$ Bohr, PM for
$9\le D \le 11.9$ Bohr, PPF at $11.9 \le D \le 12.8$ Bohr, FPF at $12.8 \le D \le 16.4$ Bohr,
PPF at $16.4 \le D \le 17.1$ Bohr, PM for $17.1 \le D \le 22.1$ Bohr, PPF at $22.1 \le D \le
25.6$ Bohr, and PM at $25.6 <D$ Bohr. Based on these results,
very thin films of Cs and Rb are expected to be ferromagnetic~\cite{okazaki1}.

In order to throw further light on this debate, it becomes crucial to understand
the electronic and magnetic structures of specific alkali metal thin films through
first-principles atomistic calculations (as opposed to the UJM). 
In particular, one needs to address the question: are thin films of alkali metals, 
in particular Cs, really ferromagnetic? For this purpose, I have done DFT calculations 
for K and Cs thin films using both LSDA and GGA. 
The main results in this brief
report are: for most thicknesses, alkali metal
thin films do not have a FM state; only a 6 layer Cs film
has a FM state, which, however, is degenerate with the 
PM state. The numerical methods employed and the results obtained are
discussed in detail in the next two sections.

\section{Method}

Calculations were performed within the framework of DFT.
VASP~\cite{vasp1,vasp2,vasp3,vasp4} was used for all the calculations. The wave functions
are expressed in a plane wave basis set with an energy
cutoff of 400 eV. The Brillouin zone integrations are
performed using the Monkhorst-Pack scheme. Ionic potentials
are represented by PAW. Both LSDA and
PW91 GGA functionals are used for the exchange-correlation energy. The preconditioned conjugate
gradient method as implemented in VASP is used for
wave function optimization, and the conjugate gradient is
used for ionic relaxation. K$(100)$ and Cs$(100)$ thin films are represented by a
repeated slab geometry. Each slab contains the desired number of $(100)$ planes
of the alkali metal. In the starting geometry, the atoms in the films were
placed at their bulk positions. Values of the bulk lattice constant used were calculated
with the particular method (LDA or GGA) in use. For the bcc bulk K, the lattice constants 
obtained were 5.04 \AA~ and 5.28 \AA~ respectively using LDA and GGA. For bcc Cs, the
corresponding values turned out to be 5.77 \AA~ and 6.13 \AA~ respectively. The 
experimental lattice constants for K and Cs are 5.23 \AA~ and 6.05 \AA~ respectively.
Consecutive slabs were
separated by a vacuum space equal to 6 atomic layers. For 1, 2 and 3 layer films, a
$(2\times 2)$ surface supercell was used. However, there were no forces
on any of the atoms along the $(100)$ planes. Therefore, for thicker films, a
$(1\times 1)$ surface supercell was used to reduce computational cost. Only one of the
atoms in the supercell was held fixed, and all other atoms were allowed relax
freely in all three directions without any symmetry constraints. As already
mentioned, there were no forces, and hence no movement of the atoms in the
$(100)$ planes. Relaxation perpendicular to the $(100)$ planes were also
rather small. A $(6\times 6\times 1)$ k-point mesh was used
for $(2\times 2)$ surface supercell, while for the $(1\times 1)$ surface
supercell, a $(8\times 8\times 1)$ k-point mesh was used. Absolute convergence
of energy with respect to energy cutoff and the number of k points was
thoroughly tested. Similar methods were used earlier to study thin Al$(110)$ films by
Ciraci and Batra~\cite{batra}.

\section{Results}

\begin{figure}[h]
\scalebox{0.35}{ \includegraphics{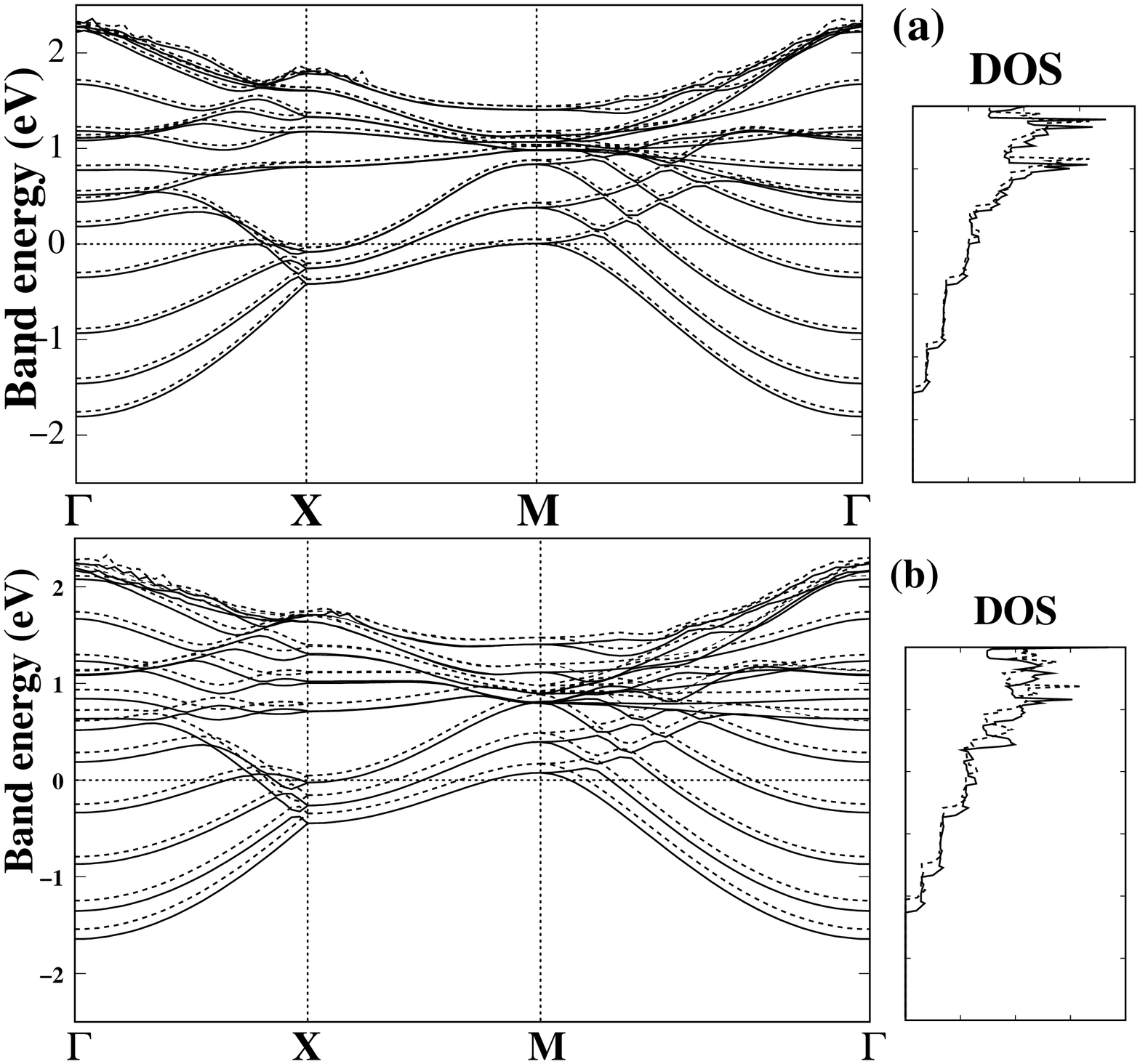} }
\hfill
 \caption{Band structure and electronic DOS for a 6 layer Cs film: (a) LSDA, and
(b) GGA results. Bands are plotted along lines joining high symmetry points in
the surface BZ of the films, and the Fermi energy has been set to zero. 
Majority (up) and minority (down) spin bands and DOS are shown
with solid and dotted lines respectively.}
 \label{fig:bandos}
 \end{figure}

Calculations were done for 3, 5 and 7 layer K films, and
for Cs films having thicknesses from 1 to 7 atomic layers. 
I searched for both paramagnetic and ferromagnetic (FM)
solutions for both K and Cs films at each thickness. I do not consider AFM states, 
since ref.~\cite{okazaki1} and ~\cite{okazaki2} do
not find an AFM ground state at any thickness.
While searching for a FM state, the initial
Hamiltonian was set up with a spin moment of 2 $\mu_B$ on each atom. In almost all cases
the films converged to PM states with zero or a vanishingly small moment
(which is probably a remanent of the initial Hamiltonian). The only exception
was a 6 layer Cs film for which a small moment of $\sim 0.1$ $\mu_B$ per atom was found 
in both LSDA and GGA. In order to be convinced that it is, in fact, a ferromagnetic
state, I plot the band structures and the electronic densities of states (DOS) in 
Figure.~\ref{fig:bandos} for a 6 layer Cs film. 
As is clearly seen, the minority (down) spin bands are slightly
shifted up in energy compared to the majority (up) spin bands. This is reflected
in the DOS as well. It should also be noted that I obtain a staircase-like DOS expected for
a quasi-two dimensional system of electrons, as discussed in detail in Ref.~\cite{batra}.
Since in all films except a 6 layer film of Cs, the initial FM films converge to a PM state,
the energies in the two states are the same. This and other results are listed in Tables~\ref{table:KLDA} 
to ~\ref{table:CsGGA}. The binding energy per atom for an $N$ atom film
given in these tables are defined as follows.
\begin{equation}
BE = \frac{1}{N}\left (N E_A-E({\rm film}) \right ),
\end{equation}
\noindent where $E_A$ is the energy of an isolated alkali metal atom, and
$E({\rm film})$ is the total energy of a film containing $N$ alkali metal atoms. 
Interestingly, even in 6 layer Cs films, the energies of the PM and FM
states are identical in LSDA, and differs by only 1 meV per atom
in GGA. This energy difference, however, is at the limit of accuracy of the present
DFT based methods. Therefore, we can only claim that a 6 layer Cs film has degenerate PM
and FM ground states.
 
\begin{table}[h]
\caption{Thickness (\AA) and binding energy (BE, eV/atom) of K$(100)$ films containing
specified number of atomic layers. Results for PM(FM) 
states are obtained using LDA(LSDA).}
\label{table:KLDA}
\begin{tabular}{|c|ccc|ccc|}
\hline
No. of layers & & PM  & &  & FM  &  \\ \cline{2-7}
   & thickness  &  & BE & thickness & & BE   \\ 
\hline
3  &  5.04  &  & 1.007   & 5.04  & & 1.007     \\
5  & 10.04  &  & 1.066   & 9.94  & & 1.066     \\
7  & 15.17  &  & 1.091   & 15.15 & & 1.091     \\
\hline
\end{tabular}
\end{table}

\begin{table}[h]
\caption{Thickness (\AA) and binding energy (BE, eV/atom) of K$(100)$ films containing
specified number of atomic layers. Results for both PM and 
FM states are obtained using PW91 GGA.}
\label{table:KGGA}
\begin{tabular}{|c|ccc|ccc|}
\hline
No. of layers & & PM  & &  & FM  &  \\ \cline{2-7}
   & thickness  &  & BE & thickness & & BE   \\ 
\hline
3  &  5.28  &  & 0.911   &  5.27  & & 0.911     \\
5  & 10.57  &  & 0.961   & 10.56  & & 0.961     \\
7  & 15.85  &  & 0.982   & 15.84  & & 0.982     \\
\hline
\end{tabular}
\end{table}

\begin{table}[h]
\caption{Thickness (\AA), binding energy (BE, eV/atom) in the PM and
FM states, and magnetization (M, $\mu_B$ per atom) 
in the FM state of Cs$(100)$ films containing 
specified number of atomic layers. Results are obtained using LDA/LSDA.}
\label{table:CsLDA}
\begin{tabular}{|c|ccc|ccc|}
\hline
No. of layers & & PM  & &  & FM  &  \\ \cline{2-7}
   & thickness  &  & BE & thickness & BE & M   \\ 
\hline
1  & $-$    &  & 0.580  & $-$    & 0.580 &  0.0   \\
2  & 3.21   &  & 0.779  & 3.18   & 0.779 &  0.0   \\
3  & 5.98   &  & 0.853  & 5.93   & 0.853 &  0.0    \\
4  & 8.93   &  & 0.884  & 8.91   & 0.884 &  0.0   \\
5  &11.56   &  & 0.908  &11.56   & 0.908 &  0.0   \\
6  &14.63   &  & 0.919  &14.61   & 0.919 &  0.1   \\
7  &17.42   &  & 0.929  &17.42   & 0.929 &  0.0    \\
\hline
\end{tabular}
\end{table}

\begin{table}[h]
\caption{Thickness (\AA), binding energy (BE, eV/atom) in the PM and
FM states, and magnetization (M, $\mu_B$ per atom) 
in the FM state of Cs$(100)$ films containing 
specified number of atomic layers. Results are obtained using PW91 GGA.}
\label{table:CsGGA}
\begin{tabular}{|c|ccc|ccc|}
\hline
No. of layers & & NM  & &  & FM  &  \\ \cline{2-7}
   & thickness  &  & BE & thickness & BE & M   \\ 
\hline
1  & $-$    &  & 0.511  & $-$    & 0.511 & 0.0  \\
2  & 3.07   &  & 0.686  & 3.07   & 0.686 & 0.0 \\
3  & 6.13   &  & 0.752  & 6.13   & 0.752 & 0.0     \\
4  & 9.25   &  & 0.774  & 9.24   & 0.774 & 0.0    \\
5  &12.30   &  & 0.794  &12.31   & 0.794 & 0.0    \\
6  &15.36   &  & 0.800  &15.33   & 0.801 & 0.1    \\
7  &18.34   &  & 0.810  &18.38   & 0.810 & 0.0     \\
\hline
\end{tabular}
\end{table}

While K-films are not expected to be ferromagnetic, as one requires thinner films for metals
with smaller $r_s$ to find a FM state according to ref.~\cite{okazaki2},
I do not find any (unique) FM ground state for Cs films either. In fact, I do not
even find a FM state except for a 6 layer Cs film. A few points are
in order here. 1. In the finite-thickness UJM that Okazaki and Teraoka have considered,
thickness is a continuous variable, but for a real metal film, it can only take discrete values
determined by the number of atomic layers in the film. 
Comparing with the LSDA results of ref.~\cite{okazaki1}, Cs films with 2 layers (thickness
of 3.21 \AA~$= 6.07$ Bohr) and 4 layers (thickness 8.93 \AA~$ = 16.88$ Bohr)
are expected to be FPF and PPF respectively. However,
both these are found to be paramagnetic in my calculations. A 6 layer film (thickness
14.63 \AA~$= 27.66$ Bohr) is found to have a FM state, albeit with a small moment,
while a finite thickness UJM is expected to be PM beyond a thickness of 25.6 Bohr.
2. Even in Okazaki and Teraoka's calculations~\cite{okazaki1,okazaki2}, 
the energy difference between a PM and a FM state
is rather small, $\sim$mev (Fig. 4 of Ref.~\cite{okazaki1}). Their calculations
are also based on the local density and local spin density approximations of the
DFT. Therefore, it would be difficult
to argue in favor of a FM ground state on the basis of those calculations, though, admittedly,
they found a definite trend at $D=7$ and $8$ with varying $r_s$. 
3. The present calculations also bring out the importance of doing an atomistic
first-principles calculation, the results of which differ significantly from those of a model
system studied in refs.~\cite{okazaki1,okazaki2}.

\section{Conclusions}

In conclusion, through first-principles atomistic calculation using
plane-wave PAW formalism within LSDA and GGA of the DFT, 
I show that neither K nor Cs films can be claimed to have
a (unique) FM ground state within the accuracy of these calculations.
Only a 6 layer Cs film has a FM ground state, but that is degenerate with a PM state.
Present work points out that there are crucial differences in the behavior of a UJM 
and a real atomistic system. More accurate calculations, such as quantum Monte Carlo,
may be required to resolve the issue of ferromagnetism in alkali metal thin films. I would like
to end by pointing out that these results, by themselves, do not rule out the possibility
of polarization of electrons in an alkali metal film in presence of TM
impurities (though that also has been argued against in refs.~\cite{kwon}
and ~\cite{sahu}), but definitely show
that within LSDA and GGA, one cannot claim alkali metal thin films to have a FM ground state.

\section{Acknowledgments}
Numerical calculations for this
study were carried out at the cluster computing facility in the Harish-Chandra
Research Institute (http://cluster.mri.ernet.in).

\vspace*{0.5in}
\centerline{\bf Figure Captions}

\end{document}